\documentclass[%
reprint,
amsmath,amssymb,
aps,
]{revtex4-2}
 
\usepackage{graphicx}% Include figure files
\usepackage{dcolumn}% Align table columns on decimal point
\usepackage{bm}% bold math
\usepackage[utf8]{inputenc}
\usepackage{physics}
\usepackage{xcolor}

\usepackage{hyperref}% add hypertext capabilities
\usepackage[mathlines]{lineno}% Enable numbering of text and display math
\begin{document}
 
\preprint{APS/123-QED}

\title{A scheme for fully programmable linear quantum networks \\based on frequency conversion}
\author{Patrick Folge}
\email[]{patrick.folge@upb.de}
\author{Michael Stefszky} % a lot of helpful discussion on interpretation an putting the work into context 
%\author{Matteo Santandrea} % Theory Help and discussion 
\author{Benjamin Brecht}
\author{Christine Silberhorn}
\affiliation{Paderborn University, Integrated Quantum Optics, Institute for Photonic Quantum Systems (PhoQS), Warburgerstr.\ 100, 33098 Paderborn, Germany}
\begin{abstract}
Linear optical quantum networks, consisting of a quantum input state and a multi-port interferometer, are an important building block for many quantum technological concepts, e.g., Gaussian boson sampling. Here, we propose the implementation of such networks based on frequency conversion by utilising a so called multi-output quantum pulse gate (mQPG). This approach allows the resource efficient and therefore scalable implementation of frequency-bin based, fully programmable interferometers in a single spatial and polarization mode. Quantum input states for this network can be provided by utilising the strong frequency entanglement of a type-0 parametric down conversion (PDC) source. Here, we develop a theoretical framework to describe linear networks based on a mQPG and PDC and utilize it to investigate the limits and scalabilty of our approach. 
\end{abstract}

\maketitle

\section{Introduction}
\label{sec:Introduction}
Linear optical quantum networks (LOQN), which we consider as a multi-port interferometer with a quantum input state and followed by photon counting or homodyne detection, have become an increasingly relevant platform and building block for many quantum technological applications. These include (Gaussian) boson sampling \cite{aaronson_computational_2013,hamilton_gaussian_2017,kruse_detailed_2019}, measurement-based quantum computation \cite{menicucci_universal_2006,gu_quantum_2009}, quantum teleportation \cite{van_loock_multipartite_2000,yonezawa_demonstration_2004}, quantum walks \cite{childs_universal_2009,venegas-andraca_quantum_2012,schreiber_photons_2010}, and quantum simulations\cite{huh_boson_2015,banchi_molecular_2020}. However, to enable useful applications of these concepts, which extend beyond proof of principle demonstrations, the underlying LOQN have to reach sufficiently high dimensionality in terms of both contributing modes and photons. Recent implementations of high dimensional LOQN were achieved in both the spatial \cite{zhong_quantum_2020} and temporal \cite{madsen_quantum_2022} degrees of freedom and were able to prove quantum computational advantages. However, these approaches require many optical components as well as synchronisation and phase stable implementation of large experimental setups. Thus, scaling these approaches is a challenging technical task. 

LOQNs can also be implemented using spectral encodings and have been explored by using electro optical modulators (EOMs) \cite{lu_quantum_2018,lu_electro-optic_2018,lu_fully_2020,kues_quantum_2019,kues_-chip_2017,lu_frequency-bin_2023} or spectrally multimode homodyne detection \cite{roslund_wavelength-multiplexed_2014,cai_multimode_2017,cai_quantum_2021}. However, the EOM based approach requires active spectral shaping of the input quantum state which can result in significant losses and the implementation of arbitrary LOQNs requires complex pulse shapes of the electrical radio frequency signals. On the other hand, the homodyne based approach faces the challenge of introducing non-Gaussian elements, which are a crucial requirement for many of the above mentioned applications, and require a phase stable implementation. 

\begin{figure}
\includegraphics[width=0.99\linewidth]{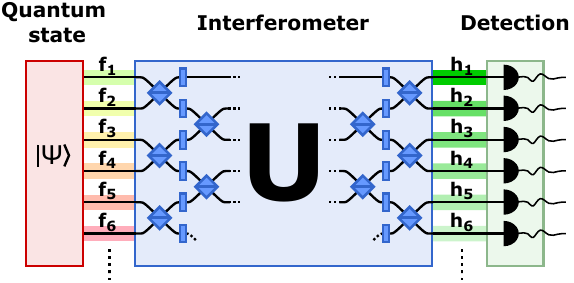}
\caption{Schematic depiction of a LOQN. The multi-port interferometer is characterized by a unitary matrix U, describing how input and output modes are connected. A quantum state is used as the quantum resource of the system.}
\label{fig:schematic_network}
\end{figure}
 
In this paper we explore an alternative approach for LOQNs in the spectral domain which is based on frequency conversion. This introduces a new platform for photonic quantum information processing and offers a highly efficient implementation of intrinsically phase stable quantum networks with full programmability. The general concept of a LOQN is depicted in Fig. \ref{fig:schematic_network} and illustrates the main requirements; controlled preparation of input quantum states, a stable but reconfigurable multi-port interferometer and  detection. At the core of our approach lies a multi-output quantum pulse gate \cite{serino_realization_2023}, allowing one to implement fully programmable frequency bin interferometer. In combination with a highly multi-mode type-0 parametric down conversion (PDC) source, one can realise a high dimensional LOQN in one spatial mode by using only two non-linear waveguides. Note, that if used together with detection in the photon number basis, our scheme does not require active phase stabilisation.

This work is organized as follows. First, we introduce the theoretical modelling of the mQPG, and discuss how it can be utilized to implement interferometers based on frequency bins. Next, we introduce type-0 PDC as an appropriate source of input quantum states for the LOQN and theoretically model the combined system of PDC and mQPG. For this we derive a formalism which allows us to investigate the quality of the frequency conversion based LOQN via the squeezing strength and purity of the output state. As an instructive example, we apply our framework to simulate a minimal example of an LOQN comprised of a frequency bin beamsplitter and squeezed input states. Finally, we investigate the fundamental limits of our scheme and explore its scalability to higher numbers of contributing modes.

\section{Theoretical model}
\label{sec:Theoretical_model}
In this work, we assume that all fields are in the form of optical pulses, which are described by a complex spectral amplitude $F(\omega)$. Such modes are usually labeled
temporal modes (TM)\cite{brecht_photon_2015}. Further, we assume for simplicity that all fields are in one spatial and polarisation mode. The creation operator of a photon in such a TM is given by \cite{brecht_photon_2015,fabre_modes_2020}
\begin{align}
\hat{F}^\dag = \int \text{d}\omega F^*(\omega) \hat{a}^\dag(\omega).    
\end{align}
We will label operators associated with a TM $F(\omega)$ with the same capital letter and a hat $\hat{F}$.

\subsection{Frequency bin Interferometer}
At the heart of a general LOQN lies a mulit-port interferometer, preferably programmable, which allows one to interfere and process the input states. 
Such an interferometer (e.g. based on spatial modes) is characterized by a unitary matrix $U_{kl}$, which describes how the (spatial) input modes $\hat{f}_l$ are connected to the (spatial) output modes $\hat{h}_k$ via the operator transformation 
\begin{align}
\hat{h}_k = \sum_{l=1}^{N_{in}} U_{kl} \hat{f}_l.   \label{eq:lin_net}
\end{align}
Here, $N_{in}$ is the number of input modes and therefore also the size of the unitary matrix. 
In other words Eq. \eqref{eq:lin_net} implies that the interferometer’s outputs
correspond to different superpositions of the inputs, while
maintaining energy conservation. 

In this work, we will present a scheme to implement such an interferometer on the basis of a set of $N_{in}$ separated frequency bins $A_l(\omega_{in})$, where $l$ labels the individual bins at central frequency  $\overline{\omega}^{in}_l$ and the $\omega_{in}$-dependence encodes the spectral profile of the bins (e.g. Gaussian). We first define a set of superposition modes 
\begin{align}
S_k(\omega_{in}) := \sum_{l=1}^{N_{in}} U_{kl} A_l(\omega_{in}), \label{eq:super_mode}
\end{align}
 which correspond to the outputs of the interferometer. The mode operators of these  then take the form $\hat{S}_k = \sum_{l=1}^{N_{in}} U_{kl}\hat{A}_l$ and contain the operators $\hat{A}_l$ pertaining to the individual bins. To implement an interferometer on the frequency bin basis, we now design a process which is capable of operating on the superposition modes $\hat{S}_k$ given by Eq. \eqref{eq:super_mode}. In the following we present the details for an experimental implementation of this task, which utilises the so called multi-output quantum pulse gate.

\subsection{The mQPG as an Interferometer}

   \begin{figure}[htp]
        \centering
        \includegraphics[width=\linewidth]{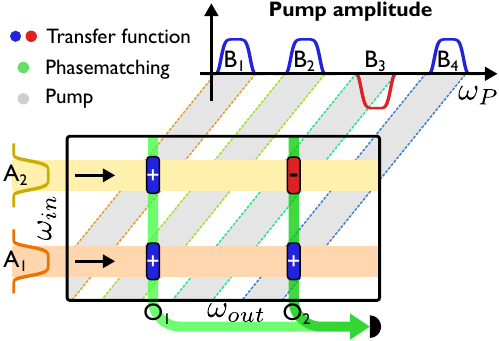}
        \caption{Schematic depiction of the transfer function of a two-output mQPG, implementing a frequency bin beam splitter.  The transfer function (red and blue) is given as the product of the phase matching function (green) and the pump spectrum (grey). Imprinting specific amplitudes and phases onto the pump allows one to program different transfer functions.}
        \label{fig:schematic_QPG}
    \end{figure}

A multi-output quantum pulse gate (mQPG) is a specially designed sum-frequency generation (SFG) process in a periodically poled non-linear waveguide \cite{brecht_demonstration_2014,serino_realization_2023}. As an SFG process, it is  characterized by a transfer function (TF)
\begin{align}
    G_{SFG}(\omega_{in},\omega_{out}) = P(\omega_P = \omega_{out} - \omega_{in}) \cdot \Phi(\omega_{in},\omega_{out})
\end{align}
which is the product of the phase-matching function $\Phi(\omega_{in},\omega_{out})$ of the nonlinear process and the complex spectrum $P(\omega_P)$ of the pump \cite{christ_theory_2013}. This TF describes how the amplitudes at input frequencies $\omega_{in}$ are converted to the output frequencies $\omega_{out}$. The distinct property of a mQPG, setting it apart from general SFG, is group velocity matching of the pump and signal fields, which can be achieved by dispersion engineering of the waveguides \cite{brecht_demonstration_2014}. Because of this, the PM-function of a mQPG is oriented perpendicular to the output-axis, which leads to a situation where the output frequency does not change for a broad input frequency range.
 
 Note, that the original quantum pulse gate \cite{brecht_demonstration_2014,eckstein_quantum_2011} had only one output, but recently the concept has been expanded for multiple outputs making it ideal for network applications \cite{serino_realization_2023}. The mQPG combines multiple spectrally separated phasematching peaks within one device, by modulating the periodic poling with a superstructure. The PM function of such an mQPG with $N_{out}$ peaks then has the form
\begin{align}
    \Phi(\omega_{in},\omega_{out}) \approx \sum_{m=1}^{N_{out}} O_m(\omega_{out}),
\end{align}
where $O_m(\omega_{out})$ describes the peak's spectral profile (typically sinc-shape) and $m$ labels the different central positions $\overline{\omega}^{out}_m$ of the peaks. The PM function of such an mQPG is depicted in Fig. \ref{fig:schematic_QPG}, where we sketch the mQPG's general working principle for two inputs and outputs. 

The mQPG allows us to perform operations on arbitrarily chosen superposition modes of frequency bins. This works under the assumption that the pump structures (here frequency bins with spectral profile $B(\omega_P)$) are spectrally broader than the individual phasematching peaks $O_m(\omega_{out})$ \footnote{This assumption ensures a single mode character of the conversion process eliminating frequency correlations}\cite{ansari_tailoring_2018}. Since the mQPG is an SFG process such a pump bin with a central frequency of $\overline{\omega}^{pump}$ addresses an input frequency bin with a central frequency of $\overline{\omega}^{in}_m = \overline{\omega}^{out}_m - \overline{\omega}^{pump}$ and converts it to the $m-$th output with a central frequency $\overline{\omega}^{out}_m$. In more detail this means that conversion is achieved at the intersection of the bins's pump function $B(\omega_P)$ and the PM function, hence, an input bin $A_{m}(\omega_{in})=B(\overline{\omega}^{out}_m-\omega_{in})$ is converted to the output mode $O_m$. Note that this input mode has the same complex spectral profile as the corresponding pump bin, but is frequency shifted. Furthermore, due to the orientation of the PM-function, the shape and position of the output modes do not change when the pump bin is shifted. This crucial feature allows for the necessary multi-path interference of interferometers, since multiple input modes can be coherently mapped to the same output 
 by utilising multiple pump bins (compare Fig. \ref{fig:schematic_QPG}).  Since the phase and amplitude of the pump bins also determines the phase and amplitude of the conversion, we can implement the mapping of one of the superposition modes $S_k$ to one of the output modes $O_m$. This is done by appropriately choosing the pump bins so that all outputs address the same input bins at centers $\overline{\omega}^{in}_l$. With this it is possible to realize a multi-port interferometer, by programming a pump spectrum of the form
\begin{align}
    P(\omega_P) = \sum_{m=1}^{N_{out}}\sum_{l=1}^{N_{in}} U_{ml} \cdot B(\overline{\omega}^{out}_m - \overline{\omega}^{in}_l - \omega_P).
\end{align}
Here, $P(\omega_P)$ is the complete pump spectrum, which is composed of individual frequency bins labeled by the corresponding frequencies of the input and output bins and weighted by the corresponding entry $U_{ml}$ of the unitary matrix describing the network. 
Using this yields a TF \begin{align}
G_{U}(\omega_{in},\omega_{out}) &= \sum_{m=1}^{N_{out}}\sum_{l=1}^{N_{in}} U_{ml} \cdot A_l(\omega_{in})\cdot O_m(\omega_{out}) \notag\\
&= \sum_{m=1}^{N_{out}} S_m(\omega_{in}) \cdot  O_m(\omega_{out}).
\label{eq:TF-multi-mode}
\end{align} 

One simple example of this scheme is depicted in Fig. \ref{fig:schematic_QPG}, namely the implementation of the TF for a balanced beamsplitter ($U_{BS} = ((1,1),(1,-1))/\sqrt{2}$) on the freqeuncy bin basis. The TF in this case is given by
\begin{align}
\begin{split}
    G_{BS}(\omega_{in},\omega_{out}) = (A_1(\omega_{in})+A_2(\omega_{in}))\cdot O_1(\omega_{out})/\sqrt{2} \\ + (A_1(\omega_{in})-A_2(\omega_{in}))\cdot O_2(\omega_{out})/\sqrt{2}. \label{eq:TF_beamsplitter}
\end{split}
\end{align}

To understand the action of such a mQPG on a quantum input state we can consider the problem within the Heisenberg picture, where a general SFG process is described via the Bogoliubov transformations \cite{christ_theory_2013}:
\begin{align}
    \hat{b}''(\omega_{in}) &= \int \text{d}\omega_{in}' \; U^Q_b(\omega_{in},\omega_{in}')\hat{b}'(\omega_{in}') \notag\\ &\qquad + \int \text{d}\omega_{out}' \; V^Q_b(\omega_{in},\omega_{out}')\hat{a}'(\omega_{out}')  \\
    \hat{a}''(\omega_{out})  &= \int \text{d}\omega_{out}' \; U^Q_a(\omega_{out},\omega_{out}')\hat{a}'(\omega_{out}') \notag\\ &\qquad - \int \text{d}\omega_{in}' \; V^Q_a(\omega_{out},\omega_{in}')\hat{b}'(\omega_{in}'). \label{eq:Bog_QPG_main}
\end{align}
Here, the operators representing the fields in front of the SFG are labeled by a single dash ($'$) and fields after the SFG by a double dash ($''$) (compare Fig. \ref{fig:schematic_network}a). We consider two different monochromatic operators $\hat{a}$ and $\hat{b}$ for input and output modes to account for the possibility of having orthogonal polarizations and for the two separated frequency ranges of $\omega_{in}$ and $\omega_{out}$. The functions $U_a^Q,V_a^Q,U_b^Q,V_b^Q$ can be calculated directly from the TF, when time ordering effects are neglected (see Appendix \ref{A4}).    Eq. \eqref{eq:Bog_QPG_main} for an mQPG with a TF \eqref{eq:TF-multi-mode} simplifies to
\begin{align}
    \hat{S}''_m = \cos(\theta_m)  \hat{S}'_m + sin(\theta_m)  \hat{O}'_m,\\
    \hat{O}''_m = \cos(\theta_m)  \hat{O}'_m - sin(\theta_m)  \hat{S}'_m.
    \label{eq:transfer_QPG}
\end{align}
These are the Heisenberg operator transformations for the superposition modes of the mQPG. The parameter $\theta_m$ defines the conversion efficiency $\sin(\theta_m)^2$ of the $m$-th mode. It can be adjusted with the pump power and can in principle reach unity \cite{reddy_high-selectivity_2018}. In this case ($\theta_m = \pi/2$) Eq. \eqref{eq:transfer_QPG} takes the form 
\begin{align}
    \hat{O}''_m = - \hat{S}'_m = -\sum_{l=1}^{N_{in}} U_{ml} \hat{A}'_i, \label{eq:fbin-inter}
\end{align}
which is equivalent to relation \eqref{eq:lin_net}, characterizing the multi-port interferometer. Note however that Eq. \eqref{eq:fbin-inter} is formulated in terms of frequency bins which are connected via frequency conversion.  

The action of a mQPG can also be interpreted as a coherent filtering of a superposition mode $S_m$ and the simultaneous quantum transduction to an output mode $O_m$. We call this process coherent filtering, because it is sensitive to the spectral phase of the considered modes. In the next section, we will describe a source of input states that are naturally compatible with the mQPG.

\subsection{Spectrally multimode squeezing source}
One desirable set of input states for LOQNs are squeezed states, for example in Gaussian boson sampling, which we consider here. An optimal source for our frequency bin based network, would deliver squeezed states in the input bins $A_k(\omega_{in})$. However, such sources are challenging to engineer and would require a sophisticated control of  the PDC process, e.g. by utilising resonators \cite{ma_highly_2023}. Therefore, we consider the use of well established degenerate type-0 PDC sources, which in the high gain regime generate squeezed states in many TMs \cite{roslund_wavelength-multiplexed_2014,kouadou_spectrally_2023}. Such PDC sources are characterized by their joint spectral amplitude (JSA)
\begin{align}
    f(\omega_{in},\omega_{in}') = P(\omega_P = \omega_{in} + \omega_{in}') \cdot \Phi(\omega_{in},\omega_{in}')
\end{align}
which is given as the product of pump amplitude spectrum and phase matching function \cite{christ_probing_2011}. Note, that since signal and idler are indistinguishable in type-0 PDC the JSA has to fulfil $ f(\omega_{in},\omega_{in}') = f(\omega_{in}',\omega_{in})$.  The evolution of an input state (here vacuum) passing through the PDC is given by the unitary operator
\begin{align}
    \hat{U}_{PDC} = \exp\left( -\frac{i}{\hbar}\int\text{d}\omega_{in}\,\text{d}\omega_{in}' f(\omega_{in},\omega_{in}') \hat{b}^\dag(\omega_{in}) \hat{b}^\dag(\omega_{in}') \right. \notag \\ \left.  + \quad \text{h.c.} \vphantom{\int_1^2} \right). \qquad  \label{eq:U_PDC}
\end{align}

\begin{figure}[htp]
        \centering
        \includegraphics[width=\linewidth]{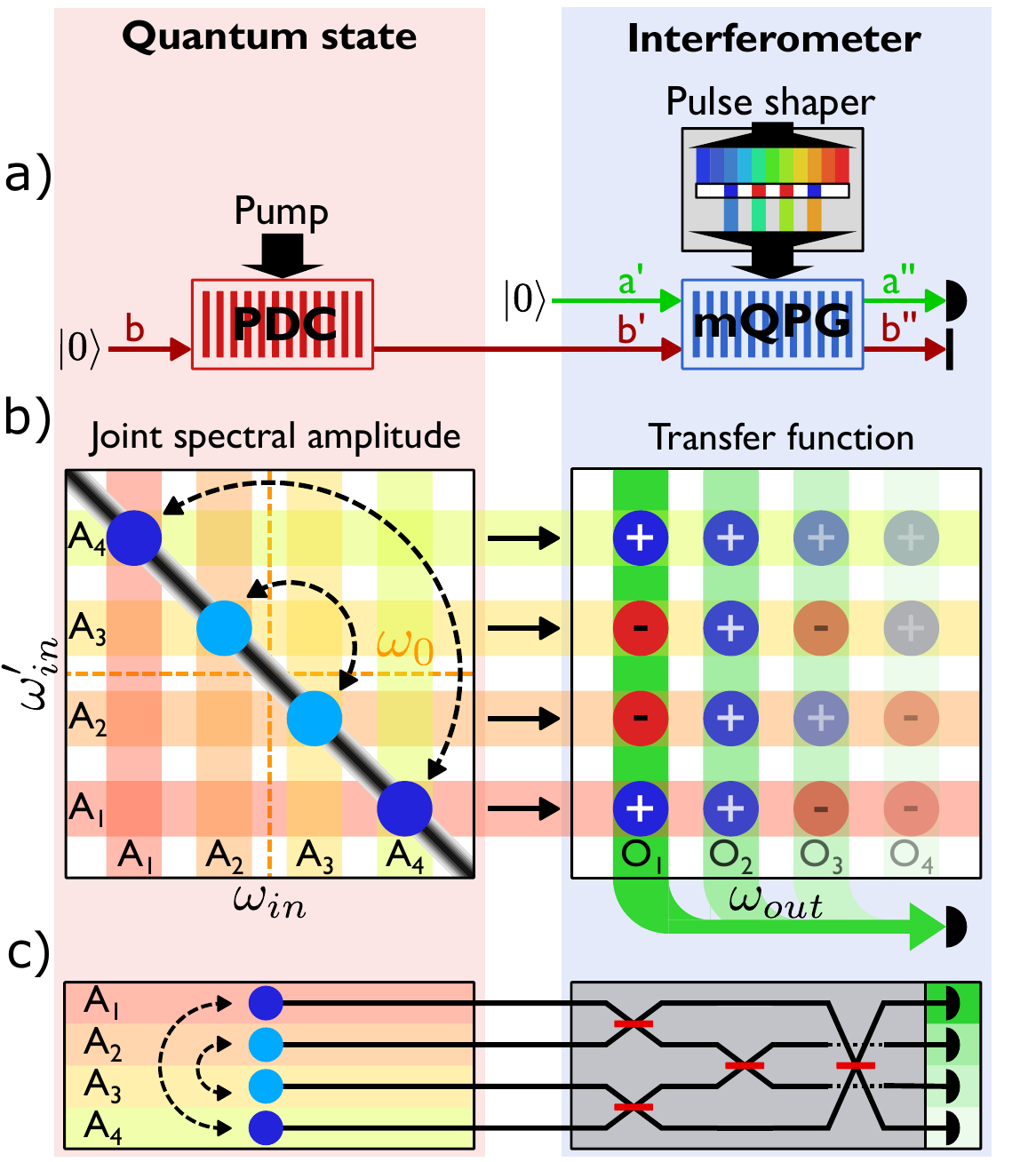}
        \caption{a) Schematic depiction of the combined system of Type-0 PDC source and mQPG. The transfer function of the mQPG can be programmed to implement an arbitrary interferometer by shaping the pump b) left: schematic depiction of the JSA in black. The blue areas highlight the effective JSA which is coherently filtered from the PDC state by the mQPG. The dashed arrows highlight different two-mode squeezed states. right: the transfer function of the mQPG which maps the coherently filtered bins into different superpositions to different output channels c) analogous interferometer in the spatial domain}
        \label{fig:schematic}
    \end{figure}

For a type-0 PDC source the JSA is given as a narrow stripe oriented along the anti-diagonal (as illustrated in Fig. \ref{fig:schematic}b). This results from the orientation of the pump function $P$ and the phase matching $\phi$ along this axis \cite{roman-rodriguez_continuous_2021}. For a very narrow  pump the JSA can be approximated by a $\delta$-function 
\begin{align}
     f(\omega_{in},\omega_{in}') &\propto \cdot\delta(\omega_{in}+\omega_{in}'-2\omega_0) \notag\\
     &\propto \sum_k \phi_k(\omega_{in}-\omega_0)\phi_k^*(-(\omega_{in}' - \omega_0)) \label{eq:narrow_PDC} 
\end{align}
which can be decomposed into  any orthonormal basis  $\{\phi_k\}$ fulfilling the  completeness relation $\delta(\omega - \omega') =\sum_k \phi_k(\omega) \phi^*_k(\omega')$. Note that in Eq. \eqref{eq:narrow_PDC} the paired functions are mirrored around the degeneracy point $\omega_0$, e.g. a bin $A_1$ at a central frequency $\omega_0 + \Delta$ is paired with a bin $A_2$ centered at  $\omega_0 - \Delta$. Since these bins are part of an orthonormal basis the unitary \eqref{eq:U_PDC} takes on the form $\hat{U}_{PDC} = \hat{U}_{12} \otimes \hat{U}_{rest}$ where the unitary describing the subspace of the bins is
\begin{align}
    \hat{U}_{12} = \exp(\alpha\hat{A}_1^\dag \hat{A}_2^\dag - \alpha^*\hat{A}_1\hat{A}_2) \label{eq:U_TMS_12}
\end{align}
and is independent of the unitary $\hat{U}_{rest}$ which describes the remaining space. Note that 
Eq. \eqref{eq:U_TMS_12} has the form of the well known two-mode squeezing (TMS) operator \cite{weedbrook_gaussian_2012}. This shows that such a PDC source provides TMS states between pairs of frequency bins. The parameter $\alpha$ combines multiple constants, including the pump strength, and determines the squeezing strength. 
However, in reality the JSAs of physical PDC sources have a finite width and the approximation of Eq. \eqref{eq:narrow_PDC} is not valid. Therefore, we consider a general description of type-0 PDC in our model, which allows us to consider any shape of the JSA. This will enable us to study the influences of it's non-negligible width in later sections. We model the PDC in the Heisenberg picture where Eq. \eqref{eq:U_PDC} takes the form of the  Bogoliubov transformation 
\begin{align}
\hat{b}'(\omega_{in}) =  \int \text{d}\omega'_{in} \; U^P(\omega_{in},\omega'_{in})\hat{b}(\omega'_{in})  \quad \quad \notag\\ + \int \text{d}\omega'_{in} \; V^P(\omega_{in},\omega'_{in})\hat{b}^{\dag}(\omega'_{in}). \label{eq:Bog_PDC_main}
\end{align}
Here, fields after the PDC are labeled with a dash (') while fields in front of the PDC do not have an additional label (compare Fig. \ref{fig:schematic}a).
Eq. \eqref{eq:Bog_PDC_main} is similar to \eqref{eq:Bog_QPG_main} of the SFG process, however only one set of monochromatic operators $\hat{b}$ is considered here, since signal and idler field have the same polarization and central frequency. The functions $U^P$ and $V^P$ can be derived from the JSA (see Appendix \ref{A3}). 

\subsection{Describing the complete LOQN}\label{sec:Combine}

In summary, our scheme to implement LOQNs reads as follows: A type-0 PDC generates TMS states between pairs of frequency bins, which are subsequently coherently filtered and superimposed in the output modes of a mQPG. The resulting quantum state in the outputs is then analogous to the output state of a spatial interferometer with TMS states in the input. In Fig. \ref{fig:schematic} we illustrate our proposed scheme for a specific example network. We depict the required experimental components  of our specific PDC source and a fully programmable mQPG. To model this combined system we adapt the theory of intensity filtered type-2 PDC presented in Ref. \cite{christ_theory_2014} to include the coherent filtering by the mQPG. This enables us to describe the frequency converted quantum state  $\rho_{out}$ in the mQPG's output in the continuous variable picture via it's covariance matrix $\sigma_{kl}$. This is possible since we consider only Gaussian transformations (squeezing and beam splitters)\cite{ferraro_gaussian_2005,braunstein_quantum_2005}. Due to the fact that the mQPG's output only consist of the modes $O_K$ we can describe the full output state on the basis of the operators $\hat{O}_k$. The quadrature operators  $\hat{X}_k = \frac{1}{\sqrt{2}}( \hat{O}_k + \hat{O}_K^\dag)$  and $\hat{Y}_k = \frac{1}{i\sqrt{2}}( \hat{O}_k - \hat{O}_k^\dag)$ corresponding to the different output modes can be arranged in the vector 
\begin{align}
    \Vec{\hat{R}} = (\hat{X}_1,\hat{Y}_1,\hat{X}_2,\hat{Y}_2,...). \label{eq:order}
\end{align}
Then the individual elements of the covariance matrix can be expressed as 
\begin{align}
\sigma_{kl}= \frac{1}{2}\left\langle\hat{R}_k \hat{R}_l + \hat{R}_l \hat{R}_k \right\rangle - \left\langle\hat{R}_k\right\rangle\left\langle\hat{R}_l\right\rangle. \label{eq:cov_mat}
\end{align}
In the following we neglect the last term because we assume vacuum states in all fields in front of the non-linear elements. Note, however, that this is not a necessity and that our framework can readily be adapted to include other input states. We describe the evolution of the states in the Heisenberg picture, by successively applying the transformations \eqref{eq:Bog_QPG_main} and \eqref{eq:Bog_PDC_main} to the operators $\hat{O}''_k$ which results in the expression
\begin{align}
    \hat{O}''_k &= \int \; \text{d}\omega_{out} \; H^1_k(\omega_{out})\hat{a}'(\omega_{out}) \notag\\
    &\qquad + \int \; \text{d}\omega_{in} \;H^2_k(\omega_{in})\hat{b}(\omega_{in})+ 
    \;H^3_k(\omega_{in})\hat{b}^\dag(\omega_{in})
\end{align}
where the amplitude functions take the form

\begin{widetext}
\begin{align}
H^1_k(\omega_{out}) &= \int \text{d}\omega'_{out} \;   O_k(\omega'_{out})U^Q_a(\omega'_{out},\omega_{out}) \notag\\
H^2_k(\omega_{in}) &= -\int \text{d}\omega'_{out} \text{d}\omega'_{in} \; O_k(\omega'_{out})V^Q_a(\omega'_{out},\omega'_{in})U^P(\omega'_{in},\omega_{in})  \notag\\
H^3_k(\omega_{in}) &= -\int \text{d}\omega_{out} \text{d}\omega'_{in} \; O_k(\omega'_{out})V^Q_a(\omega'_{out},\omega'_{in})V^P(\omega'_{in},\omega_{in}). \label{eq:amplitude_functions_Ok}
\end{align}
\end{widetext}

Inserting these operators into Eq. \eqref{eq:cov_mat} then allows one to calculate the covariance matrix for any given JSA and TF, by evaluating the vacuum expectation values. The resulting form of $\sigma_{kl}$ is derived in Appendix \ref{A5}. 
We would like to point out that our scheme, despite our description in the framework of continuous variable quantum optics, does not assume any particular detection method. Experimentally it is fully compatible with detection in the photon number basis after separating the different output channels by frequency filtering. While simulating this scenario is computationally demanding since it is effectively a GBS system, the photon number distributions can in principle be derived from the covariance matrix \cite{fitzke_simulating_2023}.
\section{Frequency beam splitter}\label{sec:F-Splitter}
    \begin{figure}[htp]
    \centering
       \includegraphics[width=\linewidth]{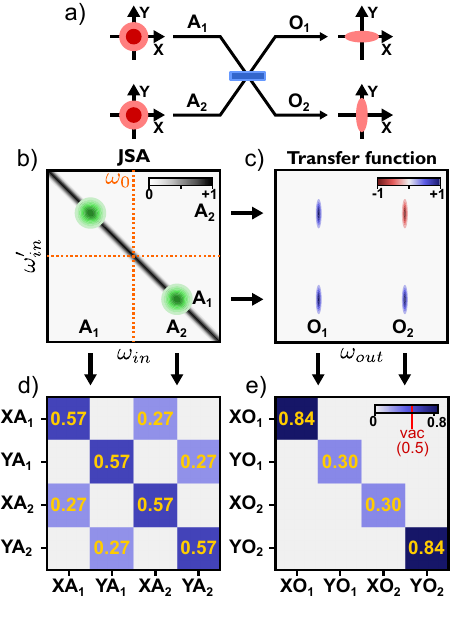}
        \caption{Simulation of a frequency beamsplitter, mapping the bins $A_1$ and $A_2$ to bins $O_1$ and $O_2$. a) Analogous spatial domain scenario b) Joint spectral amplitude (JSA) of the PDC. Green dots show the perfect two-mode squeezed JSA between bins $A_1$ and $A_2$ c) Transfer function of the mQPG. d) Absoulute value of covariance matrix between bins $A_1$ and $A_2$ after PDC, e) and between bins $O_1$ and $O_2$ after mQPG.}
        \label{fig:fsplitter}
    \end{figure}
As an instructive example of our scheme we simulate the implementation of a simple LOQN, namely the interference of both modes from a two mode squeezed state (TMS) on a balanced beamsplitter. For this we expect two independent single mode squeezed (SMS) states in the output, since this scenario is the reverse of the well known generation of TMS states by interfering SMS states on a beamsplitter \cite{weedbrook_gaussian_2012}.

The scenario is depicted in Fig. \ref{fig:fsplitter}, where we summarise the simulation by displaying the JSA and TF utilised as input for the calculation together with the resulting covariance matrices both after the PDC and at the output of the LOQN. To keep the results as general as possible we define the spectral dimensions (bin width, positions etc.) in terms of the simulation's input range $\Delta \omega_{in}$, which bounds the simulation area. In an experimental setting, this range can be understood as the bandwidth over which our scheme can operate and which is limited, for example, by the limited pump spectrum of the mQPG. To highlight the experimental feasibility of our scheme we provide simulations of realistically achievable non linear processes in periodically poled LiNbO3 waveguides in Appendix \ref{A1}, according to which we model our idealised simulations presented here. This results in a JSA which is approximated as a Gaussian cross-section of width $\text{FWHM}_{JSA} = 0.05\cdot\Delta\omega_{in}$ oriented along the anti-diagonal (compare Fig \ref{fig:fsplitter}b). We normalize this JSA to a mean photon number of $\overline{n}=1$ within the simulation region, to obtain experimentally realistic squeezing values. The frequency bin beamsplitter on the other hand is modeled by considering a TF of the form \eqref{eq:TF_beamsplitter}, where we consider Gaussian shapes for all modes ($A_1$,$A_2$,$O_1$,$O_2$). The input bins were chosen to have a width $\text{FWHM}_{bin} = 0.1\cdot\Delta\omega_{in}$, larger than $\text{FWHM}_{JSA}$.

First, we only consider the PDC and calculate the covariance matrix between two bins $A_1$ and $A_2$ which are placed symmetrically around the degeneracy point at $\omega_0$. For this we apply \eqref{eq:Bog_PDC_main} to the broadband operators $\hat{A}_1$ and $\hat{A}_2$ and then evaluate \eqref{eq:cov_mat} for the corresponding quadrature operators. As expected from the discussion above, the resulting covariance matrix (compare Fig. \ref{fig:fsplitter}c) represents a TMS state. This is evident from the sub-matrices of the individual modes, which show noise above the vacuum level of 0.5 (as one would expect from a thermal state), while being correlated when considered as as a joint system. 

The covariance matrix between the output modes $O_1$ and $O_2$ after the mQPG is derived by applying our theoretical model of the complete LOQN to discretized versions (1500x1500 points) of the JSA and TF. The resulting covariance matrix (depicted in Fig. \ref{fig:fsplitter}d) is showing two independent SMS states, which becomes apparent from the two quadrature variances (diagonal elements) which are squeezed below the vacuum level. As previously discussed, this is the expected result for the interference of a TMS state on a beamplitter and therefore establishes the capability of our scheme to implement LOQN, even when realist PDC sources with a finite JSA width are considered.

\begin{figure}[ht!]
    \centering
    \includegraphics[width=\linewidth]{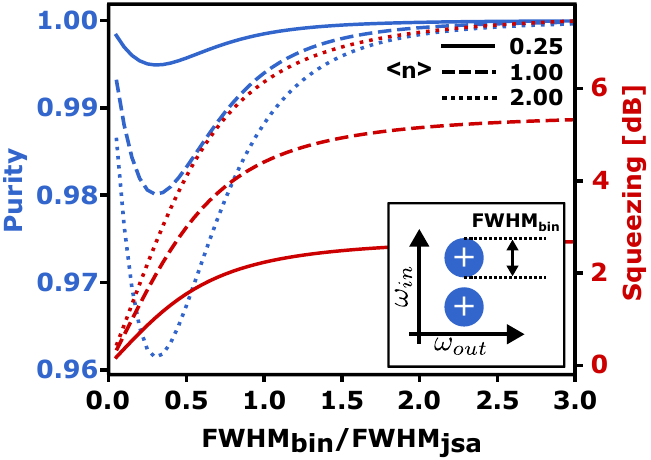}
    \caption{Squeezing and purity calculated from the covariance matrix after the freuqency beam splitter for different input bin width $\text{FWHM}_{bin}$. The 		different line types correspond to different normalizations of the JSA, which is proportional to the pump strength.}
    \label{fig:bin_scan}
\end{figure}

To better understand the limits of our scheme, we explore the quality of the output state for varying widths of the input bins $A_k$. Here, we only consider the even output ($A_1 + A_2$) of the mQPG. We quantify the quality of the output state by calculating the purity and squeezing strength of this state from the resulting covariance matrix. The purity is given by $\gamma = \text{tr}(\rho_{out}^2) = 1/(2^N\sqrt{\text{det}(\sigma)})$ \cite{ferraro_gaussian_2005} and the squeezing strength in $dB$ as $S=-10\cdot \log (2\cdot a)$ where $a$ is the  minimal eigenvalue of $\sigma$ \cite{simon_quantum-noise_1994}. We simulate these quantities for input bins in a range from $\text{FWHM}_{bin} \approx 0$ to $\text{FWHM}_{bin} = 0.15\Delta\omega_{in}$ and for three different normalizations of the JSA. These normalizations correspond to different pump strengths of the PDC process and are chosen to represent JSAs with mean photon numbers of 0.25, 1 and 2. The results are depicted in Fig. \ref{fig:bin_scan}. 
One can immediately sees from Fig. \ref{fig:bin_scan} that a minimum in purity can be observed for bins which are smaller than the width of the JSA. This can be explained by strong edge effects during the coherent filtering. Further, no clear optimal regime for operating the LOQN is observable, instead in the limit of larger bins purity and squeezing continuously improve.  This result is in contrast to heralded single photon sources from type-2 PDC, where strong spectral intensity filtering on the herald results in highly pure heralded states \cite{christ_theory_2014}, and showcases the fundamentally different behaviour of a coherent filter. 
\begin{figure*}
    \centering
    \includegraphics[width=\textwidth]{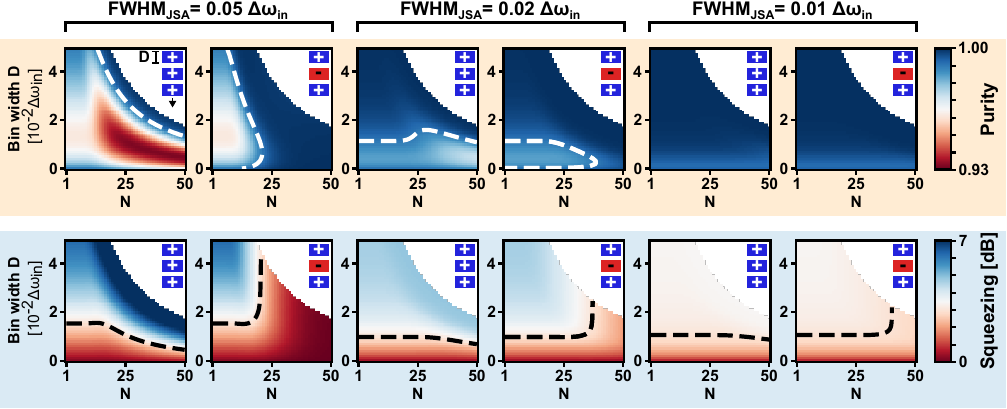}
    \caption{Investigation of squeezing an purity of the single mode squeezed state in the output channel of a mQPG after filtering from a type-0 PDC state, for varying frequency bin with and number. The white area is inaccessible, because neighboring bins overlap.  We investigate the cases of equal and alternating (0 and $\pi$) phase. Top: Purity and Bottom: Squeezing in the output channel. Three different JSAs are considered with $FWHM_{JSA} = 0.05, 0.02, 0.01 \Delta\omega_{in}$.The dashed white line corresponds to an threshold purity pf $\gamma_0 = 0.99$ and the black dashed lines correspond to a squeezing value of $S_0=3 dB$.}
    \label{fig:large_network}
\end{figure*}

\section{Scaling}\label{sec:Scaling}
We argue that our scheme is an excellent candidate for the resource efficient scaling of fully programmable LOQN to higher numbers of contributing modes, since the complete network can be achieved in only two non-linear waveguides. To understand the fundamental limits and get an estimate of achievable dimensionality of the systems we perform simulations to investigate how many bins can be implemented within the given spectral window $\Delta\omega_{in}$. For this, we consider a single output mQPG with $N$ input bins. Here, we use box shaped bins $A_k$ with a width of $D$, to make use of the complete spectral range. The bins are positioned maximally spaced, equally distributed and symmetrically placed around the degeneracy point. For the PDC we consider the JSA from the previous section, with different widths $\text{FWHM}_{JSA}$, all normalized to a mean photon number of 2. To account for different programmings of the LOQN we consider the two extremal cases of equal and alternating phases ($0$ and $\pi$) between neighboring bins for which SMS states are expected in the output. Purity and squeezing strength are depicted in Fig. \ref{fig:large_network} for varying bin widths and number of used bins. 

The upper left corner, representing big bins with sufficient separation, is expectedly the only area providing good purity and squeezing values for both cases. Therefore, the LOQN can only operate in this specific region. However, it also becomes apparent that for thinner JSAs the usable area becomes larger and more homogeneous, thereby demonstrating that the dimensionality of LOQN reachable with our approach goes well beyond the two modes of the frequency bin beamsplitter. 

We also want to highlight that the investigated widths of the JSAs are well achievable with state of the art LiNbO3 waveguides. The thinnest JSA, with $\text{FWHM}_{JSA} = 0.01 \Delta\omega_{in}$. for example well approximates the JSA achievable in a 4cm long waveguide on an input window $\Delta\omega_{in}$ corresponding to 50 nm centered at 1550 nm. In Appendix \ref{A2} we display a rough estimate of the accessible dimensionalities of our scheme. We find that, with state of the art mQPGs, input numbers in the hundreds could be expected. One limitation of our scheme is the realization of high numbers of output modes, owing to the fact that the different outputs have to share the same pump bandwidth. However, it is possible to cascade multiple mQPGs, since all superposition modes which are not addressed are passing the device unconverted. These modes can therefore be accessed by a consecutive mQPG corresponding to different outputs, albeit at the cost of increasing the number of required waveguides needed for implementation.

\section{Discussion}
The scheme presented in this work considers frequency bins as a basis for the LOQN, since these are relatively easy to shape and control. But in principle the scheme can be implemented in many other TM bases, e.g., Hermite-Gaussian modes. For these we have found similar results, with the difference that for centered HG modes the input states of the LOQN are SMS instead of TMS. Further we want to highlight, again, that our scheme does not assume any specific detection method and even the use of different detection methods in different output channels can be imagined. When only detection in the photon number basis is considered our scheme does not require any phase stability between PDC and mQPG. This is because both non-linear processes are intrinsically phase stable and a relative phase between them only results in an unknown global phase of the output modes, which is not detectable in the photon number basis. In this case two repetition rate locked pump laser sources for PDC and mQPG are sufficient for a implementation of the LOQN. 

Moreover, we want to mention that we assume perfect mQPGs (unity conversion efficiency and perfect mapping of modes) throughout this work, because we want to focus on the fundamental limits of the presented scheme. However, our theoretical framework also allows to study more complicated scenarios including imperfections, since it only considers a general TF and JSA as input. This for example allows to include multi-mode effects in the outputs of the mQPG, which can occur for imperfect PM functions. In this case one output of the mQPG is described by a bigger covariance matrix, which describes all modes contributing to said output.

\section{Conclusion}
In this work we have presented a novel scheme for the implementation of LOQNs based on frequency conversion, which utilises so-called multi-output quantum pulse gates. This approach allows one to construct fully programmable and inherently phase stable multi-port interferometer on a frequency bin basis.  We demonstrate the feasibility of this approach and its natural compatibility with broadband squeezing sources, by performing simulations based on a detailed theoretical model in the continuous variable picture.

A potential experimental implementation of LOQNs based on this approach requires only two-nonlinear waveguides for the very multi-mode input state generation and the programmable interferometer. In contrast to other encodings  (e.g. spatial or temporal domain) the achievable dimensionality of this LOQN is mainly limited by spectral shaping resolution and not by the number of utilised components (e.g. beamsplitters). Due to this, the relatively low demand on required components and  the inherent compatibility with integrated optical platforms we  believe, that this approach is a promising candidate for scaling up LOQNs towards practical applications. We find that with state-of-the art mQPGs a few hundred input modes are feasible. However, reducing the phasematching width of mQPGs, by for example utilising  resonators, could allow for much larger networks. We expect our approach to become an enabling platform for future quantum technologies thanks to its inherent scalability, full programmability, and ease of experimental implementation.

\section{Acknowledgement}
The Autors thank J. Sperling and M. Santandrea for helpful discussions.
This work was supported in part by the European Commission H2020-
FET-OPEN-RIA, (STORMYTUNE) under Grant 899587

\section{Comments}
During the preparation of the manuscript we became aware of similar work
\cite{presutti_highly_2024} 

\bibliographystyle{apsrev4-2}
\bibliography{main}

\appendix

\section{Simulated experiment}
\label{A1}
\begin{figure}[htp]
    \centering
    \includegraphics[width=0.47\textwidth]{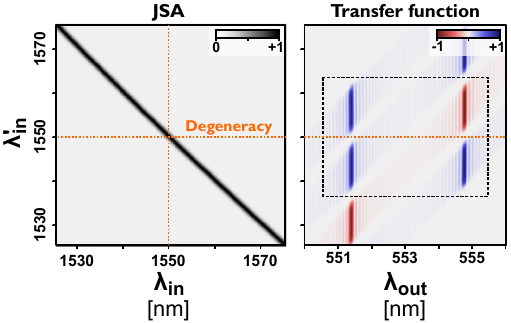}
    \caption{Simulations of left: joint spectral amplitude from a type-0 PDC process in LiNbO3 waveguide. right: the transfer function of a two-output mQPG implementing a frequency beamsplitter (represented by the dotted box)}
    \label{fig:realistic}
\end{figure}
In the main text we consider idealised systems, however to demonstrate the feasibility of the proposed systems we here provide simulations of realistically achievable nonlinear processes. These simulations are based on the Sellmeier-equations of titanium in-diffused LiNbO3 waveguides. In Fig. \ref{fig:realistic}a the joint spectral amplitude of a 1cm long waveguide, pumped with a 3 ps long pulsed laser at 775 nm, is depicted. To achieve degeneracy at 1550 nm a poling period of 16.93 $\mu$m is considered. Note, that no sinc-sidelobes are visible, since the pump width of 0.3 nm is narrower than the phasematching. For the simulation of the transfer function of a two-output mQPG (depicted in Fig. \ref{fig:realistic}b) we consider a poling period of 4.33 nm, an 1cm long waveguide and the superstructure presented in Ref. \cite{chou_multiple-channel_1999}. To simulate a frequency bin beamsplitter as discussed in the main text we consider a pump which is composed of four 3 nm wide bins. The bins are centered around a central wavelength of 860 nm and could for example be carved out from a 100 fs long pulse. Note, that these simulations utilise conservative assumptions for the design parameters, e.g. mQPG waveguides with length around 7cm are obtainable.

\section{Scalability of the Approach}
\label{A2}
Here we estimate the scalability of our approach to higher dimensions. We measure this dimensionality in terms of the number of achievable input bins $N_{in}$. This number is fundamentally limited by four factors: 1) the spectral range $\Delta \omega_{in}$  over which the type-0 PDC can provide TMS states between the frequency bins. 2) the pump bandwidth $\Delta \omega_{pump}$ of the mQPG which also limits the available input range. This bandwidth also has to be divided by the number of output bins $N_{out}$, since each output requires an equally broad pump region. 3) the phasematching width $\delta_{mQPG}$ of the mQPG because the mQPG is working under the assumption that the PM is narrower than the pump structure (bins). 4) the PM width $\delta_{PDC}$ of the PDC, since the LOQNs operation is limited by this number as discussed in the main text.

\begin{figure}
    \centering
    \includegraphics[width=0.47\textwidth]{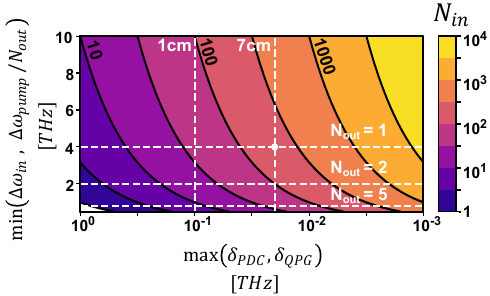}
    \caption{Estimation of the achievable number of input bins for different parameters of the available bandwidth of the network (vertical axis) and for different phasematching widths (horizontal axis). The horizontal white lines correspond to a mQPG with a pump bandwidth of 4 THz and different numbers of outputs. The vertical lines correspond to mQPGs with different lengths.}
    \label{fig:estimate_N}
\end{figure}

In this the first two points limit the available input range  while the latter two limit the minimal bin size, therefore we estimate the amount of available input bins by
\begin{align}
    N_{in}  &= \frac{\text{available input range}}{\text{minimal bin size}} \notag\\
            &= \frac{\text{min}(\Delta \omega_{in} , \Delta \omega_{pump}/N_{pump})}           {\text{max}(\delta_{PDC} , \delta_{mQPG})}.
\end{align}
The results of this estimation are depicted in Fig. \ref{fig:estimate_N}, together with the limits set by experimentally demonstrated mQPGs.\cite{serino_realization_2023,gil-lopez_improved_2021}. Considering a 7 cm long mQPG together with a 4 THz pump spectrum for example could allow for systems with 200 input bins. 

\section{Theory of Type-0 PDC}
\label{A3}
A type-0 PDC in a single spatial mode and polarization (e.g. in wavguides) can be described by the unitary operator \cite{christ_probing_2011} 
\begin{align}
    \hat{U}_{PDC} =\exp \left(  -\frac{i}{\hbar} \int\text{d}\omega_i\text{d}\omega'_i \; \text{f}(\omega_i,\omega'_i)\hat{b}^\dag(\omega_i)\hat{b}^\dag(\omega'_i) \right. \notag \\ 
    \left. \; + \; \text{h.c.} \right). \label{sup_unitary_PDC}
\end{align}
Therein, $\text{f}(\omega_i,\omega'_i)$ is the joint spectral amplitude (JSA) of the process. Here, we neglect time ordering effects, which become relevant for very strong pump fields \cite{christ_theory_2013}. In a type-0 PDC signal and idler are indistinguishable and therefore the JSA has to fulfil $f(\omega_i,\omega'_i) = f(\omega'_i,\omega_i)$. A common approach in describing PDC states is by performing a Schmidt decomposition of the JSA
\begin{align}
    -\frac{i}{\hbar}\text{f}(\omega_i,\omega'_i) &= \sum_k r_k^P \phi_k^{P*}(\omega_i) \phi_k^{P*}(\omega'_i)
\end{align}
which results in a set of orthogonal Schmidt-modes $\left\{\phi_k^P(\omega_i) \right\}$ with Schmidt-coefficients $r_k^P$ \cite{law_continuous_2000}.  These modes are equal for signal and idler because they are indistinguishable. By defining the operators $\hat{\phi}_k^\dag := \int \text{d}\omega_i \;  \phi_k^{P*}(\omega_i) \hat{b}^\dag(\omega_i)$, the Schmidt-decomposition allows  to rewrite the unitary \eqref{sup_unitary_PDC} as  
\begin{align}
    \hat{U}_{PDC} = \bigotimes_k \exp \left[ r_k^P (\hat{\phi}_k^\dag)^2  \; + \;  \text{h.c.} \right] \label{unitary_PDC_bb} = \bigotimes_k \hat{S}_k^{(SMS)}(r_k^P),
\end{align}
which corresponds to multiple independent single mode squeezing operators on the different Schmidt modes. However, besides this fundamental structure of type-0 PDC sources we show in the main text, that in the case of very multi-mode PDC, also two-mode squeezed states can be extracted from such a source.

In the Heisenberg picture, the unitary \eqref{sup_unitary_PDC} takes the form of a linear Bogoliobov transformation \cite{christ_theory_2013}:
\begin{align}
\hat{b}'(\omega_i) = \int \text{d}\omega'_i \; U^P(\omega_i,\omega'_i)\hat{b}(\omega'_i) +   \int \text{d}\omega'_i \; V^P(\omega_i,\omega'_i)\hat{b}^\dag(\omega'_i).
\label{sup_Bog_PDC}
\end{align}
Here, $U^P$ and $V^P$ can be expressed with help of the Schmidt modes $\phi_k^P(\omega_i)$ and can therefore be directly obtained from the JSA. They have the form \cite{christ_theory_2013}
\begin{align}
U^P(\omega_i,\omega'_i)  &= \sum_k \phi_k^{P*}(\omega_i) \cosh(r_k^P)\phi_k^P(\omega'_i)  \notag\\
V^P(\omega_i,\omega'_i)  &= \sum_k \phi_k^{P*}(\omega_i) \sinh(r_k^P)\phi_k^{P*}(\omega'_i).
\label{eq:UV_from_JSA}
\end{align}
\section{Theory of SFG}
\label{A4}
Because the multi-output quantum pulse gate is based on a sum frequency generation (SFG) process, it can be described by the unitary operator of a general SFG process \cite{christ_theory_2013}
\begin{align}
    \hat{U}_{SFG} = \exp \left(  -\frac{i}{\hbar} \int\text{d}\omega_{i}\text{d}\omega_{o} \; \text{G}(\omega_{i},\omega_{o})\hat{a}^\dag(\omega_o)\hat{b}(\omega_i) \right. \notag \\ \left. \; + \; \text{h.c.} \right). \label{unitary_QPG}
\end{align}
Here, $G(\omega_i,\omega_o)$ is the transfer function (TF) of the process, which describes, how the input frequencies $\omega_i$ are converted the the output frequencies $\omega_o$. Note, that we choose one of the input fields of the mQPG to be represented by the same operators $\hat{b}(\omega_i)$ as the field of the PDC process. 

In the Heisenberg picture the SFG process takes the form of the Bogoliobov transformations \cite{christ_theory_2013}
\begin{align}
    \hat{b}''(\omega_i) &= \int \text{d}\omega'_i \; U^Q_b(\omega_i,\omega'_i)\hat{b}'(\omega'_i)    \notag \\ 
    &\qquad + \int \text{d}\omega'_o \; V^Q_b(\omega_i,\omega'_o)\hat{a}'(\omega'_o) \notag \\
    \hat{a}''(\omega_o) &= \int \text{d}\omega_o' \; U^Q_a(\omega_o,\omega'_o)\hat{a}'(\omega'_o)   \notag \\ 
    &\qquad - \int \text{d}\omega'_i \;  V^Q_a(\omega_o,\omega'_i)\hat{b}'(\omega'_i).\label{Sup_Bog_QPG}
\end{align}
The functions U and V can again be calculated by performing a Schmidt decomposition of the TF which takes the form
\begin{align}
-\frac{i}{\hbar}\text{G}(\omega_i,\omega_o) = -\sum_k r_k^Q \phi_k^Q(\omega_i) \psi_k^{Q*}(\omega_o)
\end{align}
and results in the two orthonormal bases  $\left\{\phi_k^Q(\omega_i) \right\}$ and  $\left\{\psi_k^Q(\omega_o) \right\}$. This then allows to connect the Schmidt-modes to the Bogoliobov transformations via \cite{christ_theory_2013}
\begin{align}
U^Q_b(\omega_i,\omega'_i)  &= \sum_k \phi_k^{Q*}(\omega_i) \cos(r_k^Q)\phi_k^Q(\omega'_i) \notag\\
V^Q_b(\omega_i,\omega'_o) &= \sum_k \phi_k^{Q*}(\omega_i) \sin(r_k^Q)\psi_k^{Q}(\omega'_o)\notag\\
U^Q_a(\omega_o,\omega'_o)  &= \sum_k \psi_k^{Q*}(\omega_o) \cos(r_k^Q)\psi_k^Q(\omega'_o) \notag\\
V^Q_a(\omega_o,\omega'_i) &= \sum_k \psi_k^{Q*}(\omega_o) \sin(r_k^Q)\phi_k^{Q}(\omega'_i).
\label{eq:UV_from_TF}
\end{align}
Defining the broadband operators $\hat{R}_k = \int \text{d}\omega_o \psi_k^Q(\omega_o)\hat{a}(\omega_o)$ and  $\hat{H}_k = \int \text{d}\omega_i \phi_k^Q(\omega_i)\hat{b}(\omega_i)$ corresponding to the Schmidt modes allows to simply the transformation \eqref{Sup_Bog_QPG} to
\begin{align}
    \hat{H}'_k &= \cos(r_k^Q)\hat{H}_k + \sin(r_k^Q)\hat{R}_k \\
    \hat{R}'_k &=\cos(r_k^Q)\hat{R}_k - \sin(r_k^Q)\hat{H}_k.
\end{align}
These equations have the same structure as \eqref{eq:transfer_QPG}, however since we are considering general SFG the modes ($\hat{H}_k$ and $\hat{R}_k$) can spectrally overlap and are therefore not separately detectable via spectral multiplexing. This is one of the features enabled via considering a TF of form \eqref{eq:TF-multi-mode}, realizable in mQPGs, which converts to well separated output modes $O_k$. In other words, the Schmidt modes of the mQPG with a TF \eqref{eq:TF-multi-mode} are the superposition modes $S_k$ and the output modes $O_k$, with degenerate (equal weights) Schmidt coefficients.

\section{Combining PDC and mQPG}
\label{A5}
The goal of our model is the description of the output quantum state of the mQPG. Since each output channel corresponds to one mode $O_k$, this output state can be characterised in terms of the density matrix $\sigma$ on the basis of the output modes $O_k$ (compare Eq. \eqref{eq:cov_mat}). We describe the dynamics of the two non-linear processes in the Heisenberg picture by  consecutively applying \eqref{eq:Bog_PDC_main} and \eqref{eq:Bog_QPG_main} to the output operators
\begin{align}
\hat{O}''_k = \int \text{d}\omega_o O_k(\omega_o) \hat{a}''(\omega_o)
\end{align}
and obtain
\begin{align}
\hat{O}''_k &=  \int \text{d}\omega_o \; H^1_k(\omega_o) \hat{a}'(\omega_o)  \notag \\  &\qquad +  \int \text{d}\omega_i H^2_k(\omega_i) \hat{b}(\omega_i) +  H^3_k(\omega_i)\hat{b}^\dag (\omega_i) \label{sup_combined_system}
\end{align}
where we have defined the functions 
\begin{align}
H^1_k(\omega_o) &= \int \text{d}\omega'_o \;   O_k(\omega'_o)U^Q_a(\omega'_o,\omega_o) \notag\\
H^2_k(\omega_i) &= -\int \text{d}\omega'_o \text{d}\omega'_i \; O_k(\omega'_o)V^Q_a(\omega'_o,\omega'_i)U^P(\omega'_i,\omega_i)  \notag\\
H^3_k(\omega_i) &= -\int \text{d}\omega'_o \text{d}\omega'_i \; O_k(\omega'_o)V^Q_a(\omega'_o,\omega'_i)V^P(\omega'_i,\omega_i). \label{eq:O_HHH}
\end{align}
These functions can be derived from a given JSA and TF by utilising \eqref{eq:UV_from_JSA} and  \eqref{eq:UV_from_TF}. 
To now describe the output state of the mQPG, we first observe that the we can neglect displacement (second term of \eqref{eq:cov_mat}), since we assume vacuum states in front of the non-linear elements and do not consider seeding. By considering the operator order of \eqref{eq:order} the covariance matrix can be constructed from the 2x2 submatrices 
\begin{align}
\widetilde{\sigma}_{kl}=
\begin{pmatrix}
\left\langle\hat{X}_k \hat{X}_l\right\rangle + \left\langle\hat{X}_l \hat{X}_k  \right\rangle, & \left\langle \hat{X}_k \hat{Y}_l\right\rangle   + \left\langle\hat{Y}_l \hat{X}_k \right\rangle  \\
\left\langle \hat{Y}_k \hat{X}_l\right\rangle   +\left\langle \hat{X}_l \hat{Y}_k\right\rangle,  & \left\langle \hat{Y}_k \hat{Y}_l\right\rangle   + \left\langle\hat{Y}_l \hat{Y}_k\right\rangle  
\end{pmatrix},
\end{align}
where $k$ and $l$ label two modes from $\lbrace \hat{O}_k \rbrace$. The submatrices for $k=l$ describe the substates in the individual channels and for  for $k \neq l$ it describes the  the quadrature covariances between two different output modes. 
To calculate these submatrices we first express the individual elements in terms of the output operators and obtain

\begin{align}
\left\langle\hat{X}_k\hat{X}_l\right\rangle &= 
\frac{1}{2}\left\langle\ \hat{O}_k \hat{O}_l + \hat{O}_k \hat{O}_l^\dag + \hat{O}_k^\dag \hat{O}_l  + \hat{O}_k^\dag \hat{O}_l^\dag\right\rangle \notag\\
\left\langle \hat{X}_k \hat{Y}_l\right\rangle &= 
\frac{1}{2i}\left\langle\ \hat{O}_k \hat{O}_l - \hat{O}_k \hat{O}_l^\dag + \hat{O}_k^\dag \hat{O}_l  - \hat{O}_k^\dag \hat{O}_l^\dag\right\rangle \notag\\
\left\langle\hat{Y}_k \hat{X}_l\right\rangle &= 
\frac{1}{2i}\left\langle\ \hat{O}_k \hat{O}_l + \hat{O}_k \hat{O}_l^\dag - \hat{O}_k^\dag \hat{O}_l  - \hat{O}_k^\dag \hat{O}_l^\dag\right\rangle \notag\\
\left\langle\hat{Y}_k \hat{Y}_l\right\rangle &= 
\frac{-1}{2}\left\langle\ \hat{O}_k \hat{O}_l - \hat{O}_k \hat{O}_l^\dag - \hat{O}_k^\dag \hat{O}_l  + \hat{O}_k^\dag \hat{O}_l^\dag\right\rangle.  \label{eq:sup_terms_cov_mat}
\end{align}

By assuming vacuum input states and inserting \eqref{eq:O_HHH}, we are then able to calculate the terms in \eqref{eq:sup_terms_cov_mat} which results in 
\begin{align}
\bra{0} \hat{O}_k \hat{O}_l \ket{0} &=\int \text{d}\omega_i \; H^2_k(\omega_i)H^3_l(\omega_i) \\
\bra{0}  \hat{O}_k \hat{O}_l^\dag\ket{0} &= \int   \text{d}\omega_o \; H^1_k(\omega_o)H_l^{1*}(\omega_o)  \notag \\
&\qquad+ \int\text{d}\omega_i \;H^2_k(\omega_i)H_l^{2*}(\omega_i) \notag\\
\bra{0} \hat{O}_k^\dag \hat{O}_l\ket{0}  &= \int   \text{d}\omega_i \; H_k^{3*}(\omega_i) H^3_l(\omega_i) \notag\\
\bra{0}\hat{O}_k^\dag \hat{O}_l^\dag\ket{0}  &= \int   \text{d}\omega_i \; H_k^{3*}(\omega_i)H_l^{2*}(\omega_i).
\end{align}
This now allows to calculate the complete covariance matrix at the output of the mQPG. We want to mention that this approach can be applied to describe general systems comprised of a type-0 PDC and a SFG processes, since it only requires a JSA and TF as input. The output modes then take the form of the output Schmidt basis ($\psi_k^Q(\omega_{o})$) of the TF. This for example allows to study multi-mode effects occurring in imperfect mQPGs.

\end{document}